\documentclass[aps,twocolumn,showpacs]{revtex4}
\usepackage{graphicx}
\def\ee{{\rm e}}  \def\ii{{\rm i}}  \def\pb{{\bf p}}  \def\Nb{{\bf N}}
\def\rb{{\bf r}}    
\def\Eb{{\bf E}}  \def\Fb{{\bf F}}  \def\Rb{{\bf R}}  \def\Hb{{\bf H}}
\def\Lb{{\bf L}}  \def\sb{{\bf s}}  \def\nb{\hat{\bf n}}

\begin{document}

\title{Momentum transfer to small particles by aloof electron beams}

\author{F. J. Garc\'{\i}a de Abajo}
\affiliation{Centro Mixto CSIC-UPV/EHU and Donostia International
Physics Center (DIPC), Aptdo. 1072, 20080 San Sebasti\'{a}n,
Spain}
\date{\today}

\begin{abstract}
The force exerted on nanoparticles and atomic clusters by fast
passing electrons like those employed in transmission electron
microscopes are calculated and integrated over time to yield the
momentum transferred from the electrons to the particles.
Numerical results are offered for metallic and dielectric
particles of different sizes (0-500 nm in diameter) as well as for
carbon nanoclusters. Results for both linear and angular momentum
transfers are presented. For the electron beam currents commonly
employed in electron microscopes, the time-averaged forces are
shown to be comparable in magnitude to laser-induced forces in
optical tweezers. This opens up the possibility to study
optically-trapped particles inside transmission electron
microscopes.
\end{abstract}
\pacs{73.20.Mf,33.80.Ps,42.50.Vk,78.67.Bf} \maketitle


\section{Introduction}

Electromagnetic forces in optical tweezers are currently employed
to trap small particles ranging in size from nanometers to several
microns \cite{MQ00,NBX97}, and to manipulate them in all spatial
directions \cite{KTG02,G03}. This type of forces is also used to
characterize the elastic properties of deformable tiny objects
(e.g., living cells \cite{GAM00}), to obtain quantitative
information on mechanical properties at small length scales
\cite{MQ00}, and in general, to fix the position of those
particles so that they can be manipulated at will.

In this context, transmission electron microscopy offers a
potentially useful tool to study optically trapped particles,
providing excellent spatial resolution (sometimes below 1 $\AA$)
when sub-nanometer electron beams are employed \cite{BDK02}, while
allowing spectroscopic characterization with sub-eV accuracy.
Actually, transmission electron microscopes are routinely
exploited to probe local optical response properties \cite{P1985},
and more recently, also to determine photonic structures of
complex materials \cite{paper080}.

\begin{figure}
\centerline{\scalebox{0.28}{\includegraphics{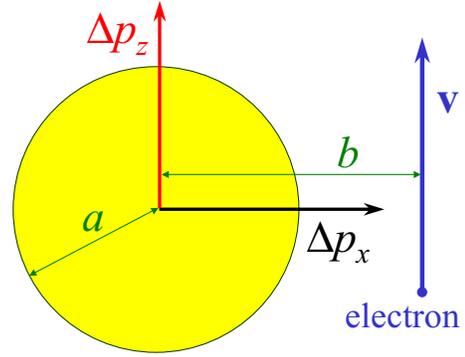}}}
\caption{(color online). Schematic representation of the process
considered in this work: a fast electron moving with impact
parameter $b$ and velocity $v$ with respect to a polarizable
particle transfers momentum $\Delta \pb=(\Delta p_x, \Delta p_z)$
to the particle via electromagnetic interaction.} \label{Fig1}
\end{figure}

A major problem that may arise when combining electron microscopy
with optical tweezers or other types of optical trapping (e.g.,
optical lattices \cite{BFG1989,BFG1990,GME02}) is that the passing
electrons can kick the particles out of the trapping locations
(see Fig.\ \ref{Fig1}). In this work, we show that the momentum
transferred from the passing electrons to the particles can be
well below the threshold needed to kick them out for commonly
employed trapping laser intensities, although a detailed
comparison between trapping forces and electron-induced forces
suggests that both weak and strong perturbation regimes are
possible depending on the distance between the particles and the
beam, all of them within the range that allows a sufficiently
large electron-particle interaction as to perform electron energy
loss spectroscopy (EELS) with significant statistics for in vacuo
optically-trapped particles.

The moving electrons can be in fact regarded as a source of
evanescent electromagnetic field that probes the sample locally,
and in this sense, they can be also used to produce deformation in
elastic particles, oscillations of trapped particles around their
equilibrium positions, and other interesting effects associated to
the transfer of momentum within accurately controlled spatial
regions.

\section{Theory}
\label{SecII}

The electromagnetic force exerted on a particle in vacuum is given
by the integral of Maxwell's stress tensor over a surface $S$
embedding the particle \cite{J1975} as
\begin{eqnarray}
  \Fb(t)&=& \frac{1}{4\pi} \int_S d\sb [\Eb(\sb,t) \, \Eb(\sb,t)\cdot\nb
                                   +\Hb(\sb,t) \, \Hb(\sb,t)\cdot\nb
  \nonumber \\
        &&\;\;\; -\frac{\nb}{2} (|\Eb(\sb,t)|^2+|\Hb(\sb,t)|^2)],
  \nonumber
\end{eqnarray}
where $\nb$ is the surface normal and Gaussian units are used. The
momentum transferred to the particle, $\Delta \pb$, is obtained by
integrating of $\Fb(t)$ over the time. This yields
\begin{eqnarray}
  \Delta \pb = \int \Fb(t) \, dt = \int_0^\infty \Fb(\omega) \, d\omega,
  \label{Dp}
\end{eqnarray}
where
\begin{eqnarray}
  \Fb(\omega)&=& \frac{1}{4\pi^2} {\rm Re} \{\int_S d\sb [\Eb(\sb,\omega) \, (\Eb(\sb,\omega)\cdot\nb)^*
  \label{eFw} \\
        && \;\;\;\;\;\;\;\;\;\;\;\; +\Hb(\sb,\omega) \, (\Hb(\sb,\omega)^*\cdot\nb)^*
  \nonumber \\
        && \;\;\; -\frac{\nb}{2} (|\Eb(\sb,\omega)|^2+|\Hb(\sb,\omega)|^2)]
        \},
  \nonumber
\end{eqnarray}
and the Fourier transform is defined as $\Eb(\rb,\omega)=\int dt
\Eb(\rb,t) \exp\{\ii\omega t\}$.

The force acting on the particle is due in part to radiation
emitted as a result of interaction with the electron and in part
to the reaction force experienced by the projectile. For small
particles, the effect of radiation emission is negligible and the
trajectory is deflected by an angle $\approx\Delta p/mv$, where
$m$ and $v$ are the mass and velocity of the electron.
Non-retarded calculations have shown that this angle is too small
to be easily measured \cite{Rivacoba}.

\subsection{Small particles}

Let us first consider a small isotropic particle sufficiently far
away from the electron beam as to neglect higher multipoles beyond
induced dipoles. The particle is then characterized by its
frequency-dependent polarizability $\alpha(\omega)$, and the force
exerted by each frequency component of the external field
$\Eb(\rb,\omega)$ reduces to \cite{EELS_force2}
\begin{eqnarray}
  \Fb(\omega)={\rm Re}\{\alpha \, \sum_j E_j^{\rm ext}(\rb,\omega) \nabla [E_j^{\rm ext}(\rb,\omega)]^*\}.
  \label{Falpha}
\end{eqnarray}
This expression can be derived from Eq.\ (\ref{eFw}) by
considering an integration surface arbitrarily close to the object
and by using the expressions for the electric and magnetic fields
induced by a small polarizable particle in terms of its
polarizability $\alpha$. For an electron moving with velocity $v$
towards to positive $z$ direction and passing by the origin at
$t=0$, the external field is readily calculated from Maxwell's
equations to yield
\begin{eqnarray}
  \Eb^{\rm ext}(\rb,\omega)=\frac{-2 e \omega}{v^2\gamma} \, \ee^{\ii \omega z/v} \,
                  [K_1(\frac{\omega R}{v\gamma})
                     \frac{\Rb}{R}
                    -\frac{\ii}{\gamma} K_0(\frac{\omega R}{v\gamma})
                    \hat{\bf z}],
  \label{Eext}
\end{eqnarray}
where $\Rb=(x,y)$ and $\gamma=1/\sqrt{1-v^2/c^2}$. Inserting Eq.\
(\ref{Eext}) into Eq.\ (\ref{Falpha}), one obtains
\begin{eqnarray}
  \Fb(\omega)= \frac{2 e^2 \omega^3}{v^5\gamma^3} [-{\rm Re}\{\alpha\}
                      \, f^\prime(\frac{\omega b}{v\gamma}) \, \hat{\bf x}
                  +2\gamma \, {\rm Im}\{\alpha\} \, f(\frac{\omega b}{v\gamma}) \,
                  \hat{\bf z}],
  \label{smallparticle}
\end{eqnarray}
where
\begin{eqnarray}
  f(\zeta)=K_1^2(\zeta)+K_0^2(\zeta)/\gamma^2,
  \nonumber
\end{eqnarray}
and the particle is taken to be situated at $\Rb=(-b,0)$ with
respect to the beam (see Fig.\ \ref{Fig1}).

Symmetry considerations lead to the conclusion that Rayleigh
scattering of the external-electron evanescent field (\ref{Eext})
produces a radiation pattern with inversion symmetry with respect
to a plane perpendicular to the trajectory. This means that the
overall transfer of momentum to the induced radiation is zero in
the small-particle limit, so that $\Delta p_z$ accounts for all
momentum transfer to the moving electron along $z$. Then, the
contribution of each $\omega$ component to the electron energy
loss rate is, within the non-recoil approximation valid for
sufficiently energetic electrons, $v F_z(\omega)$. Actually, one
finds that the identity $v F_z(\omega)=\hbar\omega P(\omega)$ is
satisfied, where $P(\omega)$ is the frequency-resolved loss
probability as previously obtained for small particles
\cite{paper041}. As a consequence, $F_z$ vanishes in the
$\omega\rightarrow 0$ limit, since $P(\omega)$ remains finite.

This behavior is quite different from $F_x$, which goes to a
finite value for small $\omega$'s, namely $F_x(\omega=0)=4 e^2
{\rm Re}\{\alpha(0)\}/v^2 b^3$. (Incidentally, momentum transfer
along $x$ produces negligible energy transfer in the non-recoil
approximation.) This latter formula can be used to derive a close
expression for $\Delta p_x$ valid for arbitrarily-large, finite
objects in the large impact parameter limit. In that case, only
small $\omega$'s contribute to $\Fb(\omega)$, due to the effective
exponential cut-off imposed by the modified Bessel functions $K_0$
and $K_1$. This means that only long wavelengths are relevant (to
which the object appears as small), so that it can be described by
its static polarizability. Then, the $\omega$ integral can be
performed numerically to yield
\begin{eqnarray}
  \Delta p_x=(5.55165 \, \gamma + \frac{1.85055}{\gamma}) \;\; \frac{e^2 {\rm Re}\{\alpha(0)\}}{v b^4}.
  \label{pxsimple}
\end{eqnarray}
For comparison, the momentum transferred to a charge $e$ at a
distance $b$ from the beam is $\Delta\pb=-(2 e^2/bv)\hat{\bf x}$.

\begin{figure}
\centerline{\scalebox{0.32}{\includegraphics{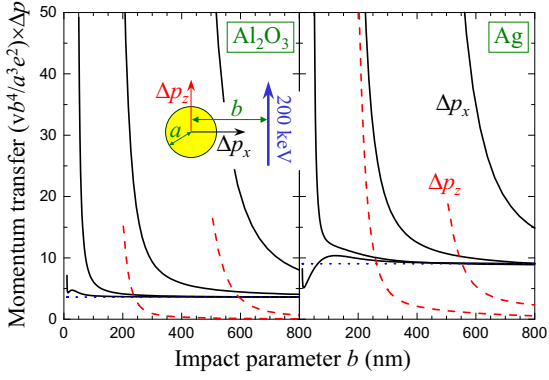}}}
\caption{(color online). Momentum transfer to small spherical
particles by a passing 200-keV electron as a function of the
distance from the trajectory to the center of the spheres $b$. The
momentum transfer has been scaled using the velocity $v=0.7 c$,
the sphere radius $a$, and the impact parameter $b$. The
perpendicular component of the momentum transfer with respect to
the trajectory $\Delta p_x$ (solid curves) has been represented
for spheres of radius $a=10$ nm, 50 nm, 200 nm, and 500 nm (notice
the rapid increase in $\Delta p_x$ near $b=a$). The parallel
component $\Delta p_z$ (dashed curves) is only shown for $a=200$
nm and 500 nm. Dielectric alumina spheres and metallic silver
spheres are considered (left and right plot, respectively),
respectively. The large $b$ limit for perpendicular momentum
transfer [Eq.\ (\ref{pxsimple})] is shown by horizontal dotted
lines.} \label{Fig2}
\end{figure}

The large-$b$ limit given by Eq.\ (\ref{pxsimple}) is compared in
Fig.\ \ref{Fig2} with more detailed calculations that include
higher-multipole moments, as described below. Also, the small
particle limit of Eq.\ (\ref{smallparticle}) is discussed in Fig.\
\ref{Fig3}.

\subsection{Arbitrary size}

For larger particles or for close electron-particle encounters,
higher multipoles become relevant in the induced forces
\cite{NRH01}. Then, it is convenient to express the evanescent
field of the electron in terms of multipoles centered at the
particle, so that the external electric and magnetic fields admit
the following decomposition \cite{paper041,paper040}:
\begin{eqnarray}
  \Eb^{\rm ext}(\rb,\omega)= \sum_{L} [\psi^{M,{\rm ext}}_{L} \Lb
                                  - \frac{\ii}{k} \psi^{E,{\rm ext}}_{L} \nabla\times\Lb]
                 j_{L}(k \rb)
  \nonumber
\end{eqnarray}
and
\begin{eqnarray}
  \Hb^{\rm ext}(\rb,\omega)= -\sum_{L} [\psi^{E,{\rm ext}}_{L} \Lb
                                  + \frac{\ii}{k} \psi^{M,{\rm ext}}_{L} \nabla\times\Lb]
                 j_{L}(k \rb),
  \nonumber
\end{eqnarray}
where $L=(l,m)$, $k=\omega/c$, $j_{L}(k\rb)=\ii^l j_l(k r)
Y_{L}(\hat{\rb})$, $\Lb=-\ii\hbar\rb\times\nabla$ is the orbital
angular momentum operator, and $\psi^{\nu,{\rm ext}}_{L}$ (for
$\nu=E,M$) are multipole coefficients given by
\cite{paper041,paper040}
   \begin{eqnarray}
      \left[ \begin{array}{c}
      \psi_{L}^{M,{\rm ext}} \\ \\
      \psi_{L}^{E,{\rm ext}}
      \end{array} \right] =
      \frac{-2\pi\ii^{1-l} e k}{l(l+1)\, \hbar c}
      \left[ \begin{array}{c}
           2 m A_{L} v/c     \\ \\
           B_{L}/\gamma
      \end{array} \right] \,
      K_m[\frac{\omega b}{v\gamma}],
   \label{eqext}
   \end{eqnarray}
with
  \begin{eqnarray}
    A_{L}=&& \sqrt{\frac{2l+1}{\pi} \frac{(l-|m|)!}{(l+|m|)!}}
           \, (2|m|-1)!!
    \nonumber \\ && \times
           \frac{\ii^{l+|m|}s_m}{(v/c)(v\gamma/c)^{|m|}}
           \,\, C^{(|m|+1/2)}_{l-|m|}(\frac{c}{v}),
    \nonumber
  \end{eqnarray}
  \begin{eqnarray}
    B_{L}= A_{l,m+1} C_+ - A_{l,m-1} C_-,
    \nonumber
  \end{eqnarray}
and
  \begin{eqnarray}
     C_\pm=\sqrt{(l\pm m+1)(l\mp m)}.
  \nonumber
  \end{eqnarray}
Here, $s_m=1$ if $m\ge 0$, $s_m=(-1)^m$ if $m<0$, and
$C^{(\nu)}_m$ is the Gegenbauer polynomial \cite{AS1972}. The
impact parameter $b$ is defined in Fig.\ \ref{Fig1}.

The induced field around the particle is given by similar
expressions obtained by substituting $\psi^{\nu,{\rm ext}}_{L}$ by
new coefficients $\psi^{\nu,{\rm ind}}_{L}$, and $j_l$ by the
Hankel function $h^{(+)}_l$ \cite{M1966}. In particular, $L=(l,m)$
is conserved for spherical particles and one has a linear
dependence $\psi^{\nu,{\rm ind}}_{L}=t_l^\nu\psi^{\nu,{\rm
ext}}_{L}$, where $t_l^\nu$ are scattering matrices that are given
by analytical expressions in the case of homogeneous particles of
dielectric function $\epsilon$ and radius $a$ \cite{paper041}:
   \begin{eqnarray}
      t_l^M = \frac{- j_l(\rho_0) \rho_1 j_l^\prime(\rho_1)
                    + \rho_0 j_l^\prime(\rho_0) j_l(\rho_1)}
                   { h_l^{(+)}(\rho_0) \rho_1 j_l^\prime(\rho_1)
                    -\rho_0  [h_l^{(+)}(\rho_0)]^\prime j_l(\rho_1)}
   \nonumber
   \end{eqnarray}
and
   \begin{eqnarray}
      t_l^E = \frac{- j_l(\rho_0) [\rho_1 j_l(\rho_1)]^\prime
                    + \epsilon [\rho_0 j_l(\rho_0)]^\prime j_l(\rho_1)}
                   { h_l^{(+)}(\rho_0) [\rho_1 j_l(\rho_1)]^\prime
                    - \epsilon [\rho_0  h_l^{(+)}(\rho_0)]^\prime j_l(\rho_1)},
   \nonumber
   \end{eqnarray}
where $\rho_0=k a$, $\rho_1=\rho_0 \sqrt{\epsilon}$ with ${\rm
Im}\{\rho_1\}>0$, and the prime denotes differentiation with
respect to $\rho_0$ and $\rho_1$.

At this point, it is convenient to write the operators $\Lb$ and
$(1/k)\nabla$ in matrix form. One finds
   \begin{eqnarray}
      \Lb j_L = \sum_{L'} \Lb_{LL'} j_{L'}
   \nonumber
   \end{eqnarray}
and
   \begin{eqnarray}
      \frac{1}{k} \nabla j_L = \sum_{L'} \Nb_{LL'} j_{L'},
   \nonumber
   \end{eqnarray}
respectively, where
   \begin{eqnarray}
\Lb_{LL'}&=&\hbar\delta_{l,l'} [C_+ \, \delta_{m+1,m'} (\hat{\bf
x}-\ii\hat{\bf y})/2
   \nonumber \\
   && \;\;\;\; + C_- \, \delta_{m-1,m'} (\hat{\bf x}+\ii\hat{\bf
y})/2 + m \, \delta_{m,m'} \, \hat{\bf z}],
   \nonumber
   \end{eqnarray}
   \begin{eqnarray}
\hat{\bf z}\cdot\Nb_{LL'}=\ii \delta_{m,m'}
(\delta_{l+1,l'}+\delta_{l-1,l'}) \frac{(l'+m)(l'-m)}{(2 l'-1)(2l'
+1)},
   \label{Nz}
   \end{eqnarray}
and the $\hat{\bf x}$ and $\hat{\bf y}$ components of $\Nb$ are
obtained from (\ref{Nz}) by rotating the reference frame using
rotation matrices for spherical harmonics \cite{M1966}. Exactly
the same matrices as above apply to $\Lb$ and $(1/k)\nabla$ acting
on Hankel functions $h_L^{(+)}$. Furthermore, these matrices
satisfy the properties $\Lb^+=\Lb$ and $\Nb^+=-\Nb$.

Now, the electric field admits an expansion of the form
  \begin{eqnarray}
    \Eb^{\rm ext}(\rb,\omega)= \sum_{L} \Eb_L^{\rm ext} j_{L}(k \rb),
  \nonumber
  \end{eqnarray}
where the coefficients
  \begin{eqnarray}
    \Eb_L^{\rm ext} = \sum_{L'} \Lb_{LL'} \psi^{M,{\rm ext}}_{L'}
    + \ii \sum_{L'L''} \Nb_{LL''}^*\times\Lb_{L''L'}^* \psi^{E,{\rm ext}}_{L'}.
  \nonumber
  \end{eqnarray}
are obtained from the above expressions. Similar formulas are
obtained for $\Hb^{\rm ext}$ and for the induced fields $\Eb^{\rm
ind}$ and $\Hb^{\rm ind}$ in terms of multipole coefficients.
Finally, we insert them into Eq.\ (\ref{eFw}) and perform the
integral over a sphere in the $s\rightarrow\infty$ limit. Then,
the first two terms inside the integrand give a vanishing
contribution because the induced far-field is transverse. The
remaining part of the integral can be recast, noticing that only
real terms must be retained,
  \begin{eqnarray}
    \Fb(\omega) &=& \frac{1}{(4\pi k)^2} \sum_{LL'} {\rm Re}\{\nb_{LL'}
\label{Fwfull} \\ &\times& (\ii\,[\Eb_L^{\rm
ext}\cdot(\Eb_{L'}^{\rm ind})^* + \Hb_L^{\rm
ext}\cdot(\Hb_{L'}^{\rm ind})^*] (1-(-1)^l) \nonumber
\\ &-& \; \ii \, [\Eb_L^{\rm ind}\cdot(\Eb_{L'}^{\rm ext})^* +
\Hb_L^{\rm ind}\cdot(\Hb_{L'}^{\rm ext})^*] (1-(-1)^{l'})
\nonumber
\\ &+& \, 2 \,[\Eb_L^{\rm ind}\cdot(\Eb_{L'}^{\rm ind})^* +
\Hb_L^{\rm ind}\cdot(\Hb_{L'}^{\rm ind})^*])\},
  \nonumber
  \end{eqnarray}
where
  \begin{eqnarray}
    \nb_{LL'} &=& \int d\Omega \, Y_{L'}^*(\Omega) \nb(\Omega) Y_L(\Omega)
  \end{eqnarray}
and $\nb(\Omega)=\sqrt{4\pi/3}\;[(\hat{\bf x}+\ii\hat{\bf
y})\,Y_{1-1}/\sqrt{2}-(\hat{\bf x}-\ii\hat{\bf
y})\,Y_{11}/\sqrt{2}+\hat{\bf z}\,Y_{10}]$ is the radial vector as
a function of the polar direction $\Omega$.

\section{Results and discussion}

Fig.\ \ref{Fig2} shows the dependence of the momentum transfer on
electron impact parameter $b$ for alumina and silver spheres of
different radius, as calculated from Eqs.\ (\ref{Dp}) and
(\ref{Fwfull}). Measured optical data have been used for the
dielectric function of these materials \cite{P1985}. One can
observe a nearly exponential decay of the momentum transfer with
$b$. Besides, the momentum transferred along the direction of the
electron velocity vector ($\Delta p_z$, dashed curves) is
generally smaller than the remaining perpendicular component
($\Delta p_x$, solid curves), which finds an explanation in the
fact that the contribution of these components to the energy loss
$\hbar\omega$ is $v\Delta p_z+(\Delta p_x)^2/m$, where $m$ is the
electron mass: since $mv\gg\Delta p$, $\Delta p_x$ is allowed to
take larger values than $\Delta p_z$ for each fixed $\omega$.

Notice also that $\Delta p_x$ converges quickly to the large $b$
limit [Eq.\ \ref{pxsimple}, dotted curves], producing a finite
result under the scaling of Fig.\ \ref{Fig2}, unlike $\Delta p_z$,
which goes faster to 0 for large $b$. In this limit, the electron
induces a dipole in the particle directed towards the electron,
which results in an attractive force between these two similar to
the image potential at surfaces \cite{jga92}, leading to a
momentum transfer $\Delta \pb \approx \Delta p_x \hat{\bf x}$. For
small metallic particles and closer encounters this picture is no
longer valid and $\Delta p_x$ can actually reverse its sign and
have a net repulsive behaviour (e.g., in Fig.\ \ref{Fig2} for Ag
particles of radius $a=10$ nm and also for the fullerenes of Fig.\
\ref{Fig4}).

\begin{figure}
\centerline{\scalebox{0.32}{\includegraphics{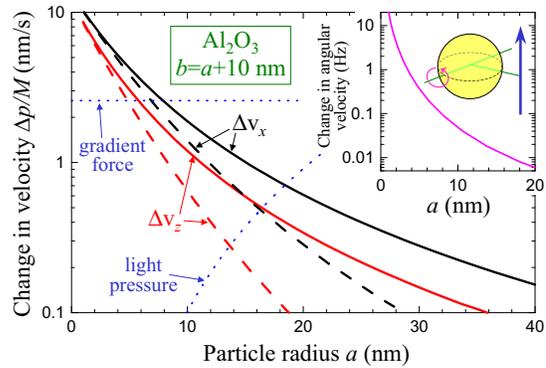}}}
\caption{(color online). Particle size dependence of the momentum
transfer normalized to the particle mass $M$ under the same
conditions as in Fig.\ \ref{Fig2}: small particle limit (dashed
curves) versus full multipole calculation (solid curves). The
particle is made of Al$_2$O$_3$ (density $\rho=4.02$ g/cm$^3$),
the electron energy is 200 keV, and the distance from the
trajectory to the particle surface is 10 nm. Dotted curves show
the momentum transferred from light in an optical trap (see text
for details). The inset depicts the change in the particle angular
velocity as a result of the torque exerted by the electron.}
\label{Fig3}
\end{figure}

\begin{figure}
\centerline{\scalebox{0.32}{\includegraphics{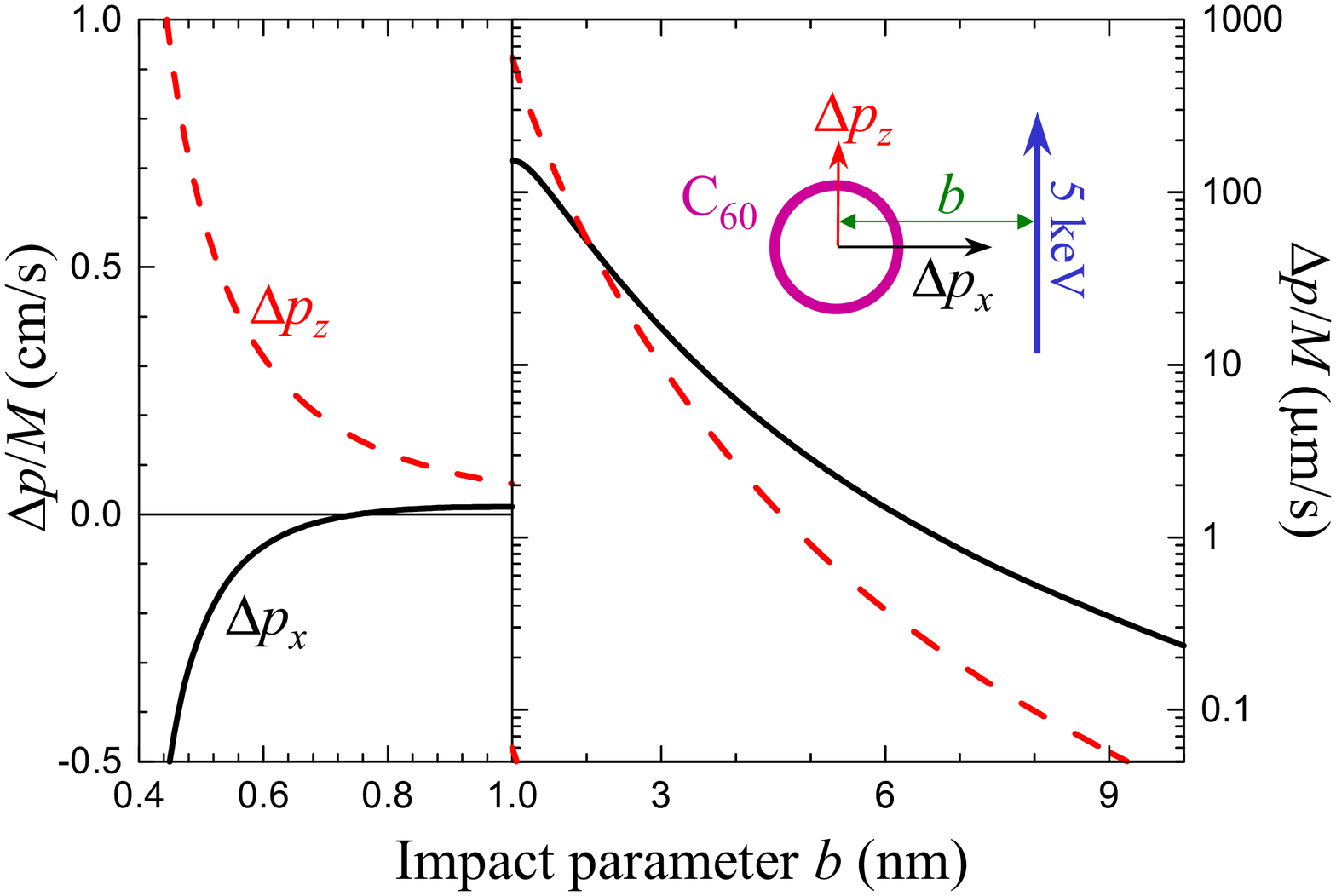}}}
\caption{(color online). Momentum transferred from a 5-keV
electron to a C$_{60}$ cluster as a function of impact parameter
$b$. The momentum is normalized to the cluster mass $M$.}
\label{Fig4}
\end{figure}

A more detailed analysis of the magnitude of the momentum transfer
effect is given in Fig.\ \ref{Fig3}. The momentum transfer is
normalized to the particle mass $M$ and the result is the change
in the particle velocity induced by the passage of the electron as
a function of particle radius $a$. The trajectory of the 200-keV
electron under consideration passes 10 nm away from the surface of
the spherical alumina particles. The full-multipole calculation
[Eqs.\ (\ref{Dp}) and (\ref{Fwfull}), solid curves] agrees well
with the small particle limit [Eqs.\ (\ref{Dp}) and
(\ref{smallparticle}), dashed curves] when $a$ is much smaller
than $b-a=10$ nm. Even though the electron-particle interaction
increases with the radius $a$, the actual change in the particle
velocity shows a nearly exponential decay with increasing $a$.

In a situation where the particle is trapped by lasers (e.g., in
optical tweezers \cite{G03} or in optical stretchers
\cite{GAM00}), one should compare the interaction with the
electrons to the interaction with the laser light. To this end, we
will consider a trapping cw-Ti:sapphire 100-mW laser emitting at a
wavelength $\lambda=785$ nm and focused on a region of radius
$R_f=10$ $\mu$m. Furthermore, we will contemplate the momentum
transferred by the laser during the average time span $\Delta t$
between two consecutive passing electrons in a transmission
electron microscope operating at a current of 1 nA. The particle
polarizability $\alpha$ is all that is needed to calculate light
forces for the small radii under discussion ($a\ll\lambda$),
according to Eq.\ (\ref{Falpha}). Now, for real $\alpha$ this
equation defines a conservative gradient force that responds to
the potential $-(\alpha/2) |\Eb|^2$, where $\Eb$ is the laser
light field, whereas the imaginary part of $\alpha$ represents
photon absorption by the particle that translates into light
pressure \cite{CT98}. These two components are represented
separately in Fig.\ \ref{Fig3} after multiplication by $\Delta
t/M$ (dotted curves). The light pressure contribution is
calculated for an incidence plane wave with the same photon flux
as the laser at its focus. The gradient force component is
obtained from the maximum force in the focus region assuming a
Gaussian profile for the laser field intensity (i.e., $|\Eb|^2
\propto \exp[-R^2/(R_f/\ln 2)^2]$). Finally, it is convenient to
define the polarizability from its relation to the scattering
matrix, which upon inspection permits writing $\alpha = (3/2 k^3)
t_1^E$. Unlike the well-known expression \cite{J1975} $\alpha=a^3
(\epsilon-1)/(\epsilon+2)$, the former relation predicts a
non-zero value for ${\rm Im}\{\alpha\}$ even for particles with
real $\epsilon$ (like our alumina spheres), arising as a pure
retardation correction associated to radiation scattering (this is
actually the origin of the light pressure component of Fig.\
\ref{Fig3}). (Incidentally, gravity would produce a velocity
change $g\Delta t=1.56$ nm/s, which is well compensated for in
currently available optical trapping systems.)

An important conclusion that can be extracted from Fig.\
\ref{Fig3} is that the crossover of trapping light into the main
source of momentum occurs for particles of 20 nm in diameter when
the electrons pass at a distance of 10 nm from the particles
surface, thus allowing one to perform energy loss analysis of the
transmitted electrons with significant statistics. Therefore,
transmission electron microscopy can be combined with in-vacuo
optical trapping to study particles of sizes above some tens nm.

While the transfer of momentum by the trapping light occurs in a
continuous smooth fashion, the electrons deposit all of the
momentum during a small time interval $\sim a/v$ ($\ll\Delta
t=0.16$ ns for 1 nA electron current). However, the change in
particle velocity per electron (vertical scale in Fig.\
\ref{Fig3}) produces a minute particle displacement during $\Delta
t$ (smaller than $1.6\times 10^{-9}$ nm $\ll a$), and therefore,
the effect of the passing electrons is experienced by the particle
as a nearly continuous source of momentum that is describable by
an average force $\Delta \pb /\Delta t$. Actually, Fig.\
\ref{Fig3} suggests that using more intense electron beams (with
even smaller impact parameters) acting during periods of the order
of one second will still not produce ejection of the particles
from their trapping locations.

It should be stressed that the momentum transfers that we have
calculated using classical electromagnetic theory must be
understood as the average value over many incoming electrons,
since the actual strength of the interaction is not large enough
as to guarantee that many photons are exchanged between each
electron and a given particle. Like in aloof EELS experiments
\cite{paper080}, most electrons will not interact with the
particles at all, so that the present results must be understood
under the perfectly valid perspective of an statistical average
performed over many beam electrons. The quadratic deviation from
these average forces can play also a role (similar to straggling
in stopping power theory), but this subject is left for future
consideration.

We have also studied momentum transfer to C$_{60}$ clusters (Fig.\
\ref{Fig4}). The scattering matrices $t^\nu_l$ have been obtained
within the discrete-dipole approximation \cite{HL96,Riva03}, where
each carbon atom is described by an induced dipole whose
polarizability is fitted to reproduce correctly the measured
optical response of graphite \cite{P1985}. Further details
concerning the procedure followed to obtain $t^\nu_l$ will be
given elsewhere \cite{gggg}. At relatively small interaction
distances $b$, the $z$ component of the momentum is larger than
the $x$ component and the latter is negative. These are effects
that can be hardly found in the above examples and that originate
in high-order multipoles (actually, $l\le 5$ are needed for
convergence within the range of $b$ under consideration). Even at
a distance of 9 nm (notice that C$_{60}$ has a diameter of only
0.7 nm) the change in velocity produced by the passing electron
can be substantial. Therefore, the interaction of fast electrons
with small clusters can produced dramatic effects if these are not
mightily bound by a mechanism stronger than optical trapping.

Finally, the passing electron can induce a torque on the particle
that changes its angular momentum ($\Delta L_y$) and makes it
rotate. This is the effect discussed in the inset of Fig.\
\ref{Fig3}, which shows the change in angular velocity per
electron, $\Delta \Omega = \Delta L_y/I$, where $I=(2/3)a^2 M$ is
the moment of inertia of the alumina sphere. Like the
electromagnetic force above, the torque is obtained from the
integral of Maxwell's stress tensor \cite{J1975}, and the details
follow a similar derivation as the one presented in Sec.\
\ref{SecII}. Averaging over the electrons of a 1 nA electron beam
passing at 10 nm from the surface of an alumina sphere of radius
$a=20$ nm, one finds an angular acceleration of 39 MHz/s. Under
these conditions, the linear momentum transferred by the electrons
can be absorbed by the trapping light, as discussed above.
However, the angular momentum is not absorbed, and the particle
will spin with increasing angular velocity until either the
centrifugal force breaks it apart or radiation emission at the
rotation frequency (vacuum friction) compensates for the
electron-induced torque.

In conclusion, we have shown that fast electrons following aloof
trajectories (i.e., without direct overlap with the sample) in a
transmission electron microscope can exert time-averaged forces on
small particles of similar magnitude as those forces associated to
trapping in optical tweezers and stretchers, and therefore, this
effect can be used for analytical studies of mechanical properties
of such particles, while electron energy loss spectra can be
actually taken without causing ejection of the particles from
their trapping positions.


\acknowledgments

The author wants to thank G. G. Hembree for suggesting this
subject and for helpful and enjoyable discussions. This work has
been partially supported by the Basque Departamento de
Educaci\'{o}n, Universidades e Investigaci\'{o}n, the University
of the Basque Country UPV/EHU (contract No. 00206.215-13639/2001),
and the Spanish Ministerio de Ciencia y Tecnolog\'{\i}a (contract
No. MAT2001-0946).

\bibliographystyle{prsty}

\end{document}